\documentclass[twocolumn,A4]{article}
\usepackage[dvips]{graphics}
\usepackage{setspace}
\usepackage{color}
\definecolor{gold}{rgb}{0.85,0.66,0}
\definecolor{dblue}{rgb}{0,0,0.5}
\begin{document}
\onecolumn
\begin{center}
{\bf{\Large {\textcolor{gold}{Quantum transport through single phenalenyl 
molecule: Effect of interface structure}}}}\\
~\\
{\textcolor{dblue}{Santanu K. Maiti}}$^{1,2,*}$ \\
~\\
{\em $^1$Theoretical Condensed Matter Physics Division,
Saha Institute of Nuclear Physics, \\
1/AF, Bidhannagar, Kolkata-700 064, India \\
$^2$Department of Physics, Narasinha Dutt College,
129, Belilious Road, Howrah-711 101, India} \\
~\\
{\bf Abstract}
\end{center}
Electronic transport characteristics through a single phenalenyl molecule 
sandwiched between two metallic electrodes are investigated by the use of 
Green's function technique. A parametric approach, based on the tight-binding 
model, is used to study the transport characteristics through such molecular 
bridge system. The electronic transport properties are significantly 
influenced by (a) the molecule-to-electrode interface structure and (b) the 
molecule-to-electrode coupling strength.
\vskip 1cm
\begin{flushleft}
{\bf PACS No.}: 73.23.-b; 73.63.Rt; 85.65.+h \\
~\\
{\bf Keywords}: Phenalenyl molecule; Conductance; $I$-$V$ characteristic.
\end{flushleft}
\vskip 5in
\noindent
{\bf ~$^*$Corresponding Author}: Santanu K. Maiti

Electronic mail: santanu.maiti@saha.ac.in
\newpage
\twocolumn

\section{Introduction}

Molecular transport have attracted much more attention since molecules 
constitute promising building blocks for future generation of electronic 
devices. Electron transport through molecules was first studied 
theoretically in $1974$ by Aviram {\em et al.}~\cite{aviram}. Since then 
numerous experiments~\cite{metz,fish,reed1,reed2,tali} have been performed 
through molecules placed between two electrodes with few nanometer 
separation. The operation of such two-terminal devices is due to an 
applied bias. Current passing across the junction is strongly nonlinear 
function of the applied bias voltage and its detailed description is a 
very complex problem. The complete knowledge of the conduction mechanism 
in this scale is not well understood even today. The transport properties 
of these systems are associated with some quantum effects, like as 
quantization of energy levels, quantum interference of electron 
waves~\cite{tagami1,mag,lau,baer1,baer2,baer3,gold,ern2}, etc. Following 
experimental developments, theory can play a major role in understanding 
the new mechanisms of conductance. The single-molecule electronics plays 
a key role in the design of future nanoelectronic circuits, but, the goal 
of developing a reliable molecular-electronics technology is still over 
the horizon and many key problems, such as device stability, 
reproducibility and the control of single-molecule transport need to 
be solved. It is very essential to control electron conduction through 
such quantum devices and the present understanding about it is quite 
limited. For example, it is not very clear how the molecular transport 
is affected by the structure of the molecule itself or by the nature of 
its coupling to the electrodes. To design molecular electronic devices 
with specific properties, structure-conductance relationships are needed 
and in a recent work Ernzerhof {\em et al.}~\cite{ern1} have presented a 
general design principle and performed several model calculations to 
demonstrate the concept.

There exist several {\em ab initio} methods for the calculation of 
conductance~\cite{tagami,yal,ven,xue,tay,der,dam} through a molecular 
bridge system. At the same time the tight-binding model has been 
extensively studied in the literature and it has also been extended to 
DFT transport calculations~\cite{elst}. The study of static density 
functional theory (DFT)~\cite{kohn} within the local-density approximation 
(LDA) to investigate the electronic transport through nanoscale conductors, 
like atomic-scale point contacts, has met with nice success. But, when this 
similar theory applies to molecular junctions, theoretical conductances 
achieve larger values compared to the experimental predictions and these 
quantitative discrepancies need extensive study in this particular field. 
In a recent work, Sai {\em et al.}~\cite{sai} have predicted a correction 
to the conductance using the time-dependent current-density functional 
theory since the dynamical effects give significant contribution in the 
electron transport, and illustrated some important results with specific 
examples. Similar dynamical effects have also been reported in some other 
recent papers~\cite{bush,ven1}, where authors have abandoned the infinite 
reservoirs, as originally introduced by Landauer, and considered two large 
but finite oppositely charged electrodes connected by a nanojunction.

Our aim of the present article is to reproduce an analytic approach based 
on the tight-binding model to characterize the electronic transport 
properties for the model of a single phenalenyl molecule and focus our 
attention on the effects of (a) the molecule-to-electrode coupling strength 
and (b) the quantum interferences in these transport. This is an interesting 
molecular system to study because of the unique behavior of the phenalenyl 
molecule and contacting the molecule to the electrodes from two different 
locations several interesting results are obtained for the electron 
transport. Here we adopt a simple parametric approach~\cite{muj1,muj2,sam,
hjo,walc1,walc2} for this calculation. The parametric study is motivated 
by the fact that the {\em ab initio} theories are computationally very 
expensive and here we concentrate only on the qualitative effects rather 
than the quantitative ones. This is why we restrict our calculations only 
on the simple analytical formulation of the transport problem. Not only 
that, the {\em ab initio} theories do not give any new qualitative behavior 
for this particular study in which we concentrate ourselves.

This paper is specifically arranged as follows. In Section $2$, we give a 
very brief description for the calculation of transmission probability ($T$), 
conductance ($g$) and current ($I$) through a finite size conductor sandwiched 
between two metallic electrodes. Section $3$ focuses the behavior of the 
conductance-energy and the current-voltage characteristics for the single 
phenalenyl molecule and studies the effects of (a) the molecule-to-electrode 
interface structure and (b) the molecular coupling strength on the above 
mentioned characteristics. These two factors i.e., the interface structure 
and the coupling strength play significant role on the quantum transport 
through molecular devices. Finally, we summarize our results in Section $4$.

\section{A glimpse onto the theoretical formulation}

Here we describe very briefly about the methodology for the calculation 
of transmission probability ($T$), conductance ($g$) and current ($I$) 
through a finite size conducting system attached to two semi-infinite 
metallic electrodes by using the Green's function technique.

\begin{figure}[ht]
{\centering \resizebox*{7.5cm}{2cm}{\includegraphics{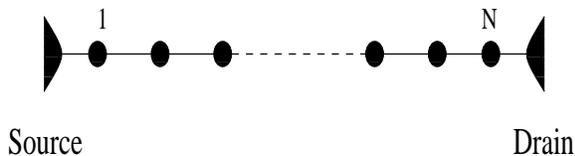}}\par}
\caption{Schematic view of a one-dimensional conductor with $N$ number of
atomic sites (filled circles) attached to two electrodes, source and drain,
through sites $1$ and $N$, respectively.}
\label{dot}
\end{figure}

Let us first consider a one-dimensional conductor with $N$ number of atomic
sites (array of filled circles) connected to two semi-infinite metallic
electrodes, namely, source and drain, as given in Fig.~\ref{dot}. The 
conducting system 
in between the two electrodes can be an array of few quantum dots, or a 
single molecule, or an array of few molecules, etc. At low voltages and 
temperatures, the conductance of the conductor can be written by using 
the Landauer conductance formula~\cite{tian,datta},
\begin{equation}
g=\frac{2e^2}{h}T
\label{land}
\end{equation}
where $g$ is the conductance and $T$ is the transmission probability of an
electron through the conductor. The transmission probability can be 
expressed in terms of the Green's function of the conductor and the 
coupling of the conductor to the two electrodes by the 
expression~\cite{tian,datta},
\begin{equation}
T={\mbox{Tr}}\left[\Gamma_S G_C^r \Gamma_D G_C^a\right]
\label{trans1}
\end{equation}
where $G_C^r$ and $G_C^a$ are respectively the retarded and advanced Green's
function of the conductor. $\Gamma_S$ and $\Gamma_D$ are the coupling terms
due to the coupling of the conductor to the source and drain, respectively. 
For the complete system i.e., the conductor including the two electrodes, 
the Green's function is defined as,
\begin{equation}
G=\left(\epsilon-H\right)^{-1}
\end{equation}
where $\epsilon=E+i\eta$. $E$ is the injecting energy of the source electron
and $\eta$ gives an infinitesimal imaginary part to $\epsilon$. Evaluation
of this Green's function requires the inversion of an infinite matrix as the
system consists of the finite conductor and the two semi-infinite electrodes.
However, the entire system can be partitioned into sub-matrices corresponding
to the individual sub-systems, and the Green's function for the conductor can
be effectively written as,
\begin{equation}
G_C=\left(\epsilon-H_C-\Sigma_S-\Sigma_D\right)^{-1}
\label{grc}
\end{equation}
where $H_C$ is the Hamiltonian for the conductor sandwiched between the two
electrodes. The single band tight-binding Hamiltonian for the conductor 
within the non-interacting picture can be written in the form,
\begin{equation}
H_C=\sum_i \epsilon_i c_i^{\dagger} c_i + \sum_{<ij>}t 
\left(c_i^{\dagger}c_j + c_j^{\dagger}c_i \right)
\label{hamil1}
\end{equation}
where $c_i^{\dagger}$ ($c_i$) is the creation (annihilation) operator of an
electron at site $i$, $\epsilon_i$'s are the site energies and $t$ is the
nearest-neighbor hopping integral. In Eq.~\ref{grc}, $\Sigma_S=h_{SC}^{\dagger}
g_S h_{SC}$ and $\Sigma_D=h_{DC} g_D h_{DC}^{\dagger}$ are the self-energy
operators due to the two electrodes, where $g_S$ and $g_D$ are respectively
the Green's function for the source and the drain. $h_{SC}$ and $h_{DC}$ are 
the coupling matrices and they will be non-zero only for the adjacent points 
in the conductor, $1$ and $N$ as shown in Fig.~\ref{dot}, and the electrodes 
respectively. The coupling terms $\Gamma_S$ and $\Gamma_D$ for the conductor 
can be calculated through the expression,
\begin{equation}
\Gamma_{\{S,D\}}=i\left[\Sigma_{\{S,D\}}^r-\Sigma_{\{S,D\}}^a\right]
\end{equation}
where $\Sigma_{\{S,D\}}^r$ and $\Sigma_{\{S,D\}}^a$ are the retarded and
advanced self-energies, respectively, and they are conjugate with each
other. Datta {\em et al.}~\cite{tian} have shown that the self-energies
can be expressed like as,
\begin{equation}
\Sigma_{\{S,D\}}^r=\Lambda_{\{S,D\}}-i \Delta_{\{S,D\}}
\end{equation}
where $\Lambda_{\{S,D\}}$ are the real parts of the self-energies which
correspond to the shift of the energy eigenvalues of the conductor and the
imaginary parts $\Delta_{\{S,D\}}$ of the self-energies represent the
broadening of these energy levels. This broadening is much larger than the
thermal broadening and this is why we restrict our all calculations in this
article only at absolute zero temperature. The real and imaginary parts of 
the self-energies can be determined in terms of the hopping integral
($\tau_{\{S,D\}}$) between the boundary site of the conductor and the
electrodes, the injection energy ($E$) of the transmitting electron and the
hopping strength ($v$) between nearest-neighbor sites of the electrodes.

Thus the coupling terms $\Gamma_S$ and $\Gamma_D$ can be written in terms of 
the retarded self-energy as,
\begin{equation}
\Gamma_{\{S,D\}}=-2{\mbox{Im}} \left[\Sigma_{\{S,D\}}^r\right]
\end{equation}
Now all the information regarding the conductor to electrodes coupling are
included into the two self energies as stated above and are analyzed through
the use of Newns-Anderson chemisorption theory~\cite{muj1,muj2}. The detailed
description of this theory is obtained in these two references.

Hence, by calculating the self-energies, the coupling terms $\Gamma_S$ and
$\Gamma_D$ can be easily obtained and then the transmission probability ($T$)
will be computed from the expression as mentioned in Eq.~\ref{trans1}.

Since the coupling matrices $h_{SC}$ and $h_{DC}$ are non-zero only for
the adjacent points in the conductor, $1$ and $N$ as shown in Fig.~\ref{dot},
the transmission probability becomes,
\begin{equation}
T(E,V)=4 \Delta_{11}^S(E,V) \Delta_{NN}^D(E,V)|G_{1N}(E,V)|^2
\label{trans2}
\end{equation}
where $\Delta_{11}=<1|\Delta|1>$, $\Delta_{NN}=<N|\Delta|N>$ and
$G_{1N}=<1|G_C|N>$.

The current passing through the conductor is depicted as a single-electron
scattering process between the two reservoirs of charge carriers. The
current-voltage relation is evaluated from the following
expression~\cite{datta},
\begin{equation}
I(V)=\frac{e}{\pi \hbar}\int \limits_{E_F-eV/2}^{E_F+eV/2} T(E,V) dE
\end{equation}
where $E_F$ is the equilibrium Fermi energy. For the sake of simplicity, 
here we assume that the entire voltage is dropped across the 
conductor-electrode interfaces and this assumption does not greatly change 
the qualitative behaviors of the $I$-$V$ characteristics. This assumption 
is based on the fact that the electric field inside the molecule, especially 
for short molecules, seems to have a minimal effect on the 
conductance-voltage characteristics. On the other hand for quite longer 
molecules and high bias voltage, the electric field inside the molecule 
may play a more significant role depending on the internal structure of 
the molecule~\cite{tian}, yet the effect is too small. Using the expression 
of $T(E,V)$ as in Eq.~\ref{trans2} the final form of $I(V)$ will be,
\begin{eqnarray}
I(V) &=& \frac{4e}{\pi \hbar}\int \limits_{E_F-eV/2}^{E_F+eV/2}
\Delta_{11}^S(E,V) \Delta_{NN}^D(E,V) \nonumber \\
& & \times |G_{1N}(E,V)|^2 dE
\label{curr}
\end{eqnarray}
Eqs.~\ref{land}, \ref{trans2} and \ref{curr} are the final working
formule for the calculation of conductance $g$, transmission probability 
$T$, and current $I$, respectively through any finite size conductor 
sandwiched between two electrodes.

By using the above methodology, in this article we will investigate the 
electronic transport characteristics through a single phenalenyl molecule 
(Fig.~\ref{phenalenyl}). Throughout this article we set $E_F=0$ and choose
the unit $c=e=h=1$.

\section{Results and their interpretation}

\begin{figure}[ht]
{\centering \resizebox*{6cm}{3.25cm}{\includegraphics{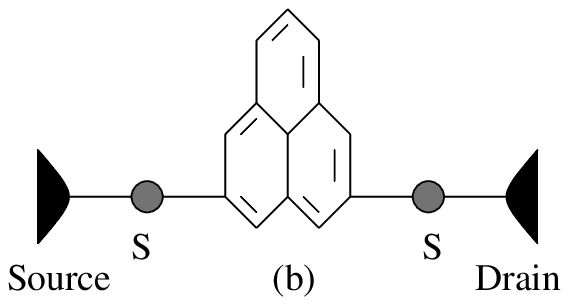}}\par}
\caption{Schematic view of single phenalenyl molecules attached to two 
metallic electrodes, namely, source and drain, through thiol (S-H) groups.}
\label{phenalenyl}
\end{figure}

In this section we focus on the conductance-energy and current-voltage 
characteristics of a single phenalenyl molecule and investigate the 
dependence of these characteristics on (a) the molecule-to-electrode 
interface structure and (b) the molecular coupling strength. The schematic 
representations of single phenalenyl molecules attached to the two metallic 
electrodes are shown in Fig.~\ref{phenalenyl}. To
\begin{figure}[ht]
{\centering \resizebox*{6cm}{3.25cm}{\includegraphics{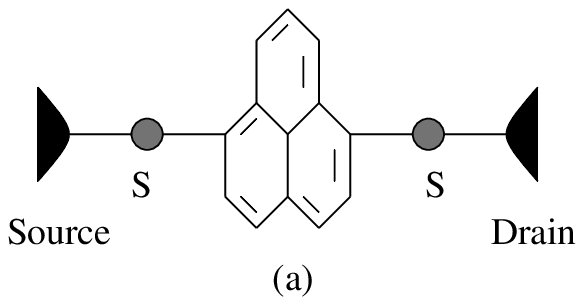}}\par}
\end{figure}
characterize the molecule-to-electrode interface effect, here we consider 
two different geometries (see Figs.~\ref{phenalenyl}(a) and (b)). These 
single molecules are connected to the electrodes by thiol (S-H) groups. In 
experiments, two electrodes made from gold (Au) are used and molecules 
attached to the electrodes by thiol (S-H) groups in the chemisorption 
technique where hydrogen (H) atoms remove and sulfur (S) atoms reside. 
Here the molecule is described by the similar tight-binding Hamiltonian 
as prescribed in Eq.~\ref{hamil1}. Throughout the article we describe 
all the essential features of electron transport in two distinct regimes. 
One is $\tau_{\{S,D\}} << t$, called the 
\begin{figure}[ht]
{\centering \resizebox*{8cm}{9cm}{\includegraphics{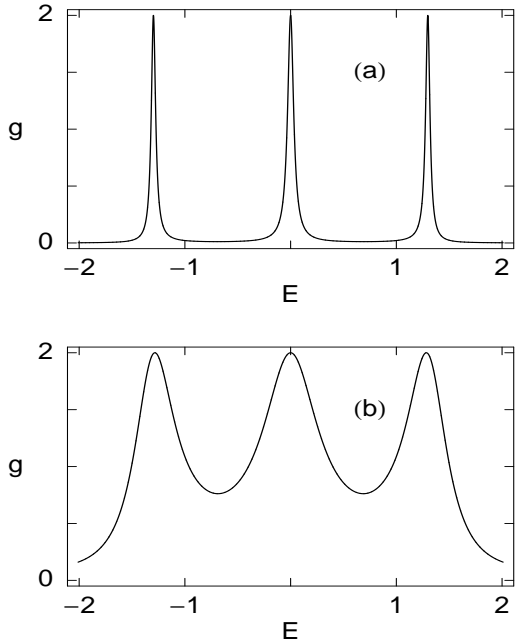}}\par}
\caption{Conductance $g$ as a function of the energy $E$ for the phenalenyl
molecule connected to the electrodes according to Fig.~\ref{phenalenyl}(a). 
(a) and (b) correspond to the results for the weak and strong 
molecule-to-electrode coupling limits, respectively.}
\label{condmol1}
\end{figure}
weak-coupling limit and the other one is $\tau_{\{S,D\}} \sim t$, called the 
strong-coupling limit, where $\tau_S$ and $\tau_D$ are the hopping strengths 
of the molecule to the source and drain, respectively. The common set of 
values of the parameters used in our calculations for these two limiting 
cases are: $\tau_S=\tau_D=0.5$, $t=2.5$ (weak-coupling) and 
$\tau_S=\tau_D=1.5$, $t=2.5$ (strong-coupling). We set the nearest-neighbor 
hopping strength $v=4$.

Figure~\ref{condmol1} shows the variation of the conductance $g$ as a 
function of the injecting electron energy $E$ for the molecular bridge 
given in Fig.~\ref{phenalenyl}(a). Figure~\ref{condmol1}(a) corresponds 
to the result for the bridge system in the weak coupling limit. The 
conductance is almost everywhere zero, except at the resonances where 
it approaches to $2$. The resonant peaks in the conductance spectrum 
coincide with the energy eigenvalues of the single phenalenyl molecule. 
Therefore the conductance spectrum manifests itself the electronic 
structure of the molecule. On the other hand, in the strong coupling 
limit the resonant peaks get substantial widths, as shown in 
Fig.~\ref{condmol1}(b) and it emphasizes that the electron conduction 
takes place throughout the energy range (the range taken here in the 
figure). This is due to the broadening 
\begin{figure}[ht]
{\centering \resizebox*{8cm}{9cm}{\includegraphics{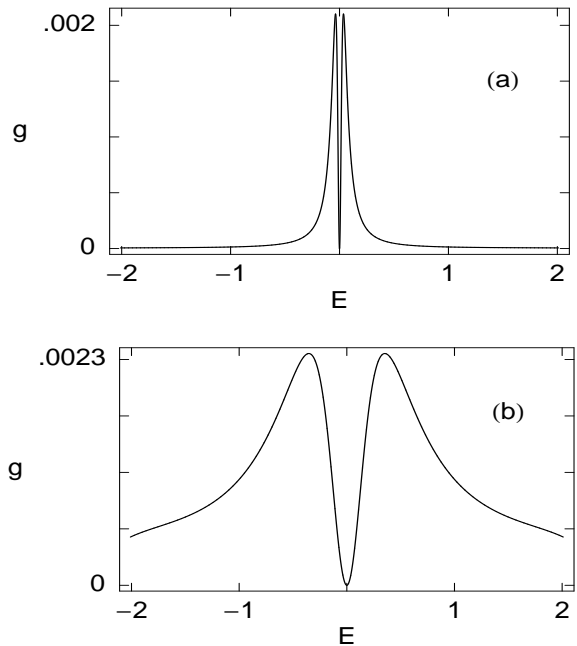}}\par}
\caption{Conductance $g$ as a function of the energy $E$ for the phenalenyl 
molecule connected to the electrodes according to Fig.~\ref{phenalenyl}(b). 
(a) and (b) correspond to the results for the weak and strong 
molecule-to-electrode coupling limits, respectively.}
\label{condmol2}
\end{figure}
of the molecular energy levels in the strong coupling case, where the
contribution comes from the imaginary parts of the 
self-energies~\cite{tian}, as described earlier.

To characterize the molecule-to-electrode interface effect on electron
transport, we plot the conductance in Fig.~\ref{condmol2} for the molecular
bridge given in Fig.~\ref{phenalenyl}(b). Figures~\ref{condmol2}(a) and (b)
correspond to the results for the bridge system in the weak and strong 
coupling limits, respectively. The broadening of the conductance peaks in 
the strong coupling limit (see Fig.~\ref{condmol2}(b)) is due to the same 
reason as mentioned earlier. From the curves plotted in Fig.~\ref{condmol2}, 
we see that the conductance peaks do not reach to unity anymore and get 
much reduced value, compared to the results described in Fig.~\ref{condmol1}. 
Such a behavior can be understood as follows. The electrons are carried 
from the source to drain through the molecule and thus the electron waves 
propagating along the two arms of the molecular ring may suffer a phase 
shift between themselves, according to the result of quantum interference 
between the various pathways that the electron can take. Therefore, the 
probability amplitude of the electron across the molecular ring becomes 
strengthened or weakened (from the standard interpretation of the wave 
functions). It emphasizes itself especially as transmittance cancellations 
and anti-resonances in the transmission (conductance) spectrum. Here the 
mentioned phase shift is observed by the variation of the geometry of the 
molecular bridge. The most significant issue for considering these two
different molecular bridge systems is that in this way the interference 
conditions are changed quite significantly.

Another key feature observed from the conductance spectrum is the existence 
of the conductance (transmittance) zero. From Fig.~\ref{condmol2}, it is 
observed that the conductance zero appears across $E=0$. Such anti-resonant 
state is specific to the interferometric nature of the scattering and does 
not occur in usual one-dimensional scattering problems of potential barriers. 
It is also observed that the position of the anti-resonant state on the 
energy scale is independent of the molecule-to-electrode coupling strength. 
The width of this state is very small and hence it does not give any 
significant contribution to the current-voltage characteristics. However, 
the variations of the interference conditions have strong influence on the 
magnitude of the current flowing through the bridge systems.

Thus it can be emphasized that the electron transmission is strongly 
influenced by the molecule-to-electrode interface structure. 

\begin{figure}[ht]
{\centering \resizebox*{8cm}{9cm}{\includegraphics{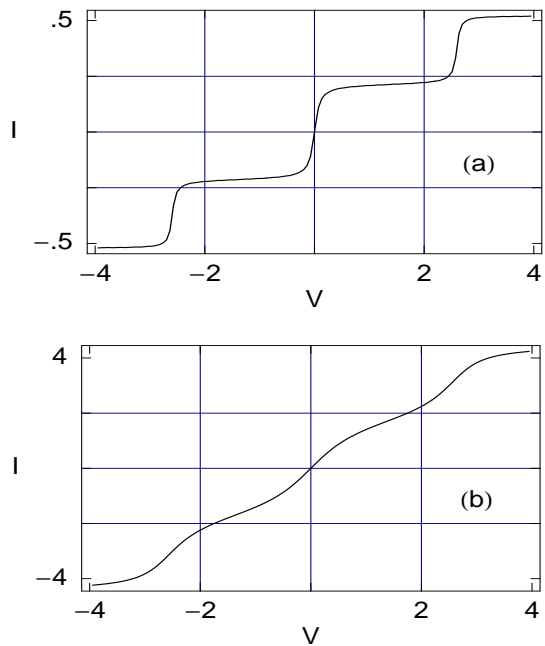}}\par}
\caption{Current $I$ as a function of the applied bias voltage $V$ for the 
phenalenyl molecule connected to the electrodes according to
Fig.~\ref{phenalenyl}(a). (a) and (b) correspond to the results for the 
weak and strong molecule-to-electrode coupling limits, respectively.}
\label{currmol1}
\end{figure}

The scenario of electron transfer through the molecular junction is much
more clearly visible from the current-voltage characteristics. Current
through the molecular system is computed by the integration procedure 
(given in Eq.~\ref{curr})~\cite{datta} of the transmission function $T$. 
The behavior of the transmission function is similar to that of the 
conductance variation since $g=2T$ (from the Landauer conductance formula, 
Eq.~\ref{land}, with $e=h=1$ in our present formulation). In 
Fig.~\ref{currmol1}, we plot the current-voltage characteristics for the 
molecular bridge shown in Fig.~\ref{phenalenyl}(a). Figures~\ref{currmol1}(a) 
and (b) correspond to the currents for the bridge system in the weak and 
strong coupling cases, respectively. It is observed that in the weak 
coupling case the current shows staircase-like structure with sharp steps. 
This is due to the discreteness of molecular resonances as shown in 
Fig.~\ref{condmol1}(a). As the voltage increases, the electrochemical 
potentials on the electrodes are shifted and eventually cross one of the 
molecular energy levels. Accordingly, a current channel is opened up and 
a jump in the $I$-$V$ curve appears. The shape and height of these current 
steps depend on the width of the molecular resonances. With the increase 
of the molecule-to-electrode coupling strength, the current varies 
continuously with the applied bias voltage and gains much more bigger 
values, as shown in Fig.~\ref{currmol1}(b). This behavior can be clearly
observed by noting the area under the curve of the conductance spectrum given
in Fig.~\ref{condmol1}(b). Comparing the results plotted in 
Figs.~\ref{currmol1}(a) and (b) it is clearly observed that the current 
amplitude gets an order of magnitude enhancement with the increase of the 
molecular coupling strength. So the electron transport through the molecular 
bridge is significantly affected by the molecule-to-electrode coupling 
strength.

\begin{figure}[ht]
{\centering \resizebox*{8cm}{9cm}{\includegraphics{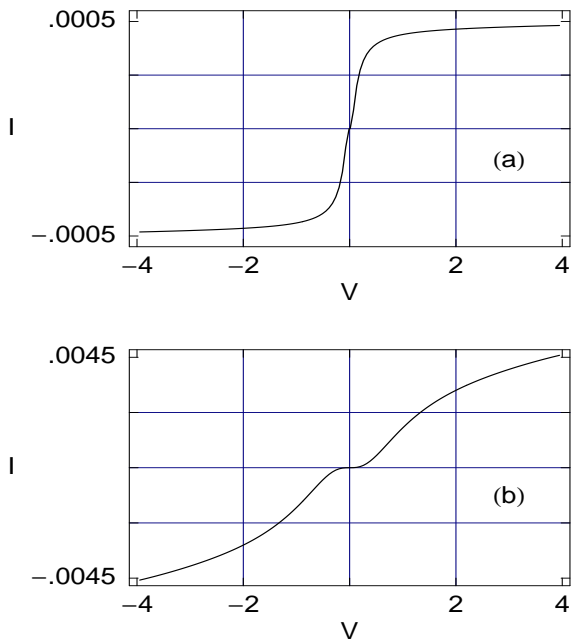}}\par}
\caption{Current $I$ as a function of the applied bias voltage $V$ for the 
phenalenyl molecule connected to the electrodes according to  
Fig.~\ref{phenalenyl}(b). (a) and (b) correspond to the results for the 
weak and strong molecule-to-electrode coupling limits, respectively.}
\label{currmol2}
\end{figure}

The effect of the molecule-to-electrode interface structure on the electron
transport through the phenalenyl molecule is much more clearly explained 
from Fig.~\ref{currmol2}, where we plot the currents for the molecular bridge  
given in Fig.~\ref{phenalenyl}(b). Figures~\ref{currmol2}(a) and (b) 
correspond to the currents in the two limiting cases as in 
Fig.~\ref{currmol1}. Similar to the previous case, here also the current 
amplitude gets an order of magnitude enhancement with the increase of the 
molecular coupling strength. 
But the significant observation is that for this bridge 
(Fig.~\ref{phenalenyl}(b)) the current amplitudes get reduced enormously 
compared to the results predicted for the molecular bridge given in 
Fig.~\ref{phenalenyl}(a) (see the results given in Fig.~\ref{currmol1}). 
This is solely due to the quantum interference effect between the different 
pathways that the electron can take. So it can be emphasized that designing 
a molecular device is strongly influenced by the molecule-to-electrode 
interface structure.
 
All the above mentioned characteristics are also valid if the 
electron-electron interaction is taken into account. The main effect of 
the electron correlation is to shift and to split the resonant positions. 
This is due to the fact that the on-site Coulomb repulsive energy $U$ gives 
a renormalization of the site energies. Depending on the strength of the 
nearest-neighbor hopping integral ($t$) compared to the on-site Coulomb 
interaction ($U$) different regimes appear. For the case $t/U<<1$, the 
resonances and anti-resonances would split into two distinct narrow bands 
separated by the on-site Coulomb energy. On the other hand, for the case 
where $t/U>>1$, the resonances and anti-resonances would occur in pairs.

\section{Concluding remarks}

To summarize, a parametric approach based on the tight-binding model has 
been used to investigate the electronic transport characteristics of a 
single phenalenyl molecule sandwiched between two metallic electrodes. 
Here we have used the Green's function method to determine the transmission 
probability ($T$), the conductance ($g$) and the current-voltage ($I$-$V$) 
characteristics through the molecule. This technique can be used to study the 
electronic transport in any complicated molecular bridge system. This is an 
interesting molecular system to study because of the unique behavior of the 
phenalenyl molecule. In this article, contacting the phenalenyl molecule 
from two different locations leads to a large change in magnitude and shape 
in the transmission spectrum and the associated current-voltage curves. 
Electron conduction through the molecule is strongly influenced by (a) the 
molecule-to-electrode coupling strength and (b) the interface structure. 
These findings indicate that designing a whole system that includes not 
only molecule but also molecule-to-electrode coupling and interface 
structure are highly important in fabricating molecular electronic devices.

More studies are expected to take the Schottky effect, comes from the charge 
transfer across the metal-molecule interfaces, the static Stark effect, which 
is taken into account for the modification of the electronic structure of the 
molecular bridge due to the applied bias voltage (essential especially for 
higher voltages). Here we have also ignored the effects of inelastic 
scattering processes and electron-electron correlation to characterize the 
electron transport through such molecular bridge systems.

\end{document}